\documentclass[a4paper,11pt]{article}

\usepackage{multicol}
\usepackage{graphicx}
\usepackage{setspace}
\usepackage[
   pdftitle={ Xe ${\rm 4d}$ photoionization in Xe@C$_{60}$, Xe@C$_{240}$, and Xe@C$_{60}$@C$_{240}$}, 
   pdfauthor={V. K. Dolmatov and D. A. Keating},         
   pdfkeywords={photoionization, correlation, endoheadral atoms},   
   colorlinks,
   linkcolor=blue,
   urlcolor=blue,
   citecolor=blue,
   a4paper=true,
dvipdfm           
   ]{hyperref}

\newcommand{\abstracttitle}[1]{
 \begin{center}{\Large {\bf #1}}\end{center}
}
\newcommand{\authors}[1]{
 \vspace*{-0.3cm}
 \begin{center} {\bf #1} \end{center}
 \vspace*{-0.3cm}
}
\newcommand{\addresses}[1]{
 \begin{center} {\small #1} \end{center}
}
\newcommand{\synopsis}[1]{
 \begin{center}
 \setstretch{0.75}
 \begin{minipage}[t]{16cm}
   {\footnotesize {\bf Synopsis} #1 }
 \end{minipage}
 \setstretch{1.0}
 \end{center}
}
\newcommand{\abstracttext}[1]{
 \vspace*{-0.3cm}
 \columnsep0.75cm
 \begin{multicols}{2} #1 \end{multicols}
}

\newcommand{\picturelandscape}[2]{
 \vspace*{0.5cm}
 \centerline{
  \includegraphics*[width=7.8cm,angle=#1]{#2}
 }
}
\newcommand{\capt}[2]{
 \vspace*{-0.3cm}
 \begin{center}
 \begin{minipage}[t]{7.8cm} {\small {\bf Figure~#1}.~#2} \end{minipage}
 \end{center}
 \vspace*{0.3cm}
}

\newcommand{\writeto}[1]{
 \hspace*{-2.5mm} \footnote{E-mail: \href{mailto:#1}{#1}}\hspace*{-1.5mm}
}

\begin{document}

\abstracttitle{Xe ${\rm 4d}$ photoionization in Xe@C$_{60}$, Xe@C$_{240}$, and Xe@C$_{60}$@C$_{240}$
}

\authors{
V. K. Dolmatov\writeto{vkdolmatov@una.edu}
 and D. A. Keating
}

\addresses{Department of Physics and Earth Science, University of North Alabama, Florence, AL 35632, U.S.A.}

\synopsis{Re-evaluated parameters for the square-well potential model for photoionization of endo-fullerenes are proposed and employed to reveal the
spectacular modifications in the Xe ${\rm 4d}$ photoionization giant resonance along the path from Xe@C$_{60}$ to Xe@C$_{240}$ to multi-walled Xe@C$_{60}$@C$_{240}$.}

\abstracttext{Resonances, termed \textit{confinement resonances} (CR's), in photoionization of atoms $A$ confined inside the C$_{60}$ cage --  A@C$_{60}$ endo-fullerenes --
have been the subject of numerous theoretical studies \cite{AQC09} (and references therein). There, many important discoveries have been made, in particular, in the frameworks the $\Delta$-potential model
\cite{AQC09}, which approximates the C$_{60}$ cage by the square-well potential of a certain finite width $\Delta$, inner radius $r_{0}$, and
 depth $U_{0}$, and the $\delta$-potential model \cite{Amusia} which assumes the C$_{60}$ cage to be infinitesimally thin. Both models
 have an indisputable value in providing a transparent initial understanding
which aspects of the  photoionization spectrum of an encapsulated atom $A$ are most unusual, thereby identifying the most useful measurements which could be performed.
Thus, theoretical predictions of a spectacular splitting of the Xe@C$_{60}$ ${\rm 4d}$ photoionization giant resonance by CR's have served an impetus for the recent
measurement of this phenomenon in Xe@C$_{60}^{+}$ \cite{Phaneuf2010}. Victoriously, it has confirmed a general prediction of the existence of CR's in endo-fullerene spectra.
The experiment has also revealed noticeably stronger developed CR's in the Xe ${\rm 4d}$ giant resonance than those predicted in the
$\Delta$-potential model \cite{AQC09}. It is argued in the present paper that while the size of C$_{60}$
 and its binding strength are somewhat robust parameters, the thickness $\Delta$ of the C$_{60}$ cage is less robust and can be empirically
 varied until a closer match of calculated and experimental results is achieved. We thus find $\Delta^{\rm new} \approx 1.25$,
 $r_{0}^{\rm new} \approx 6.01$, and $U_{0}^{\rm new} \approx -0.422$\ {\em au} compared to the formerly used  $\Delta^{\rm old} \approx 5.8$, $r_{0}^{\rm old} \approx 5.8$,
and $U_{0}^{\rm old} \approx -0.302$\ {\em au} \cite{AQC09}.
 The calculated data for the Xe ${\rm 4d}$  photoionization cross section $\sigma_{\rm 4d}$, upon the Xe@C$_{60}^{+}$ photoionization, obtained with the use of both the new and old $\Delta$, $r_{0}$
and $U_{0}$, are compared with experiment \cite{Phaneuf2010} in figure $1,a$.
A much closer agreement between experiment and the thus calculated new $\sigma_{\rm 4d}$ ($\sigma_{\rm 4d}^{\rm new}$) (solid line) is obvious.

 To advance the study of CR's in the Xe ${\rm 4d}$ photoionization giant resonance in endo-fullerenes, we have extended it to the giant Xe@C$_{240}$
 and nested Xe@C$_{60}$@C$_{240}$ endo-fullerenes; the latter was accounted for in the nested-$\Delta$-potential model developed in \cite{Onions}.
  With the above found thickness $\Delta \approx 1.25$\ {\em au} in mind,
 we find  that, for C$_{240}$,  $r_{0} \approx 12.875$ and $U_{0} \approx -0.52$\ {\em au}.
The calculated results are depicted in figure $1,b$ which unravels impressive modifications in the emerging CR's along the path
Xe@C$_{60}$ $\rightarrow$ Xe@C$_{240}$ $\rightarrow$ Xe@C$_{60}$@C$_{240}$. We challenge experimentallists to observe them.

 \picturelandscape{0}{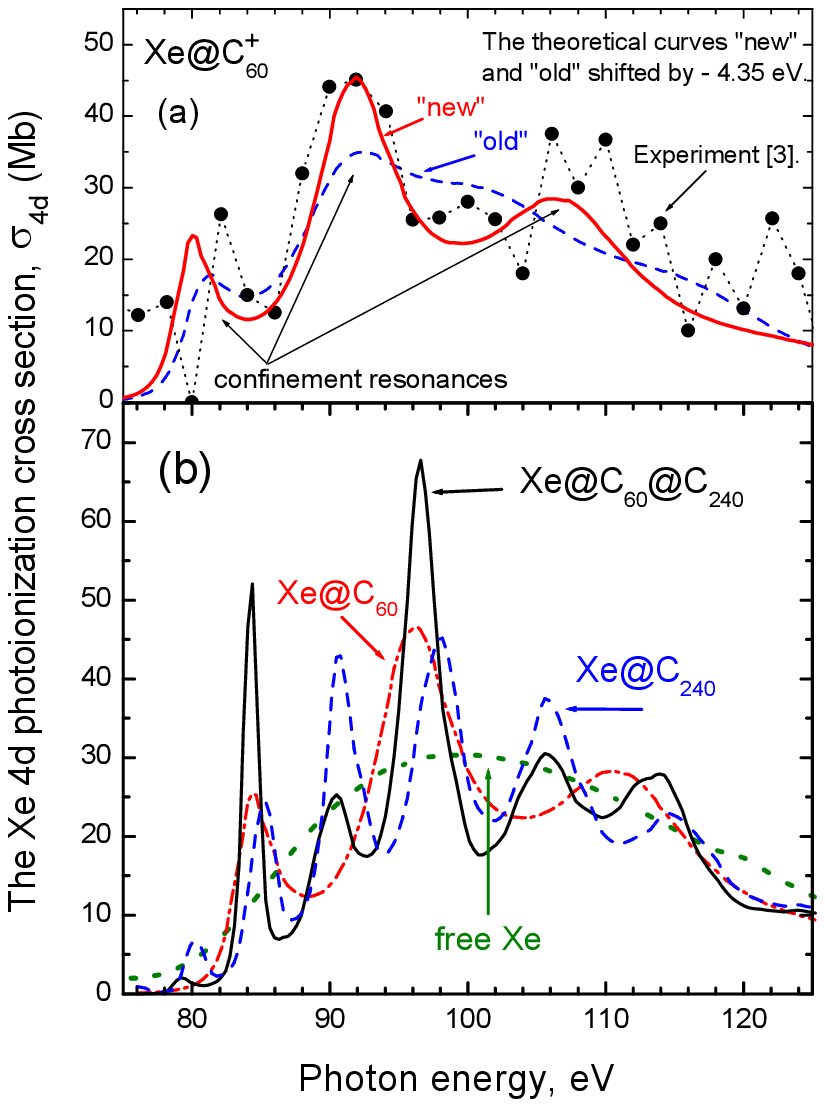}
\capt{1}{The Xe ${\rm 4d}$ photoionization cross section $\sigma_{\rm 4d}$ in the Xe-endo-fullerenes and free Xe, as marked;
 $\sigma_{4d}^{\rm exp}$ \cite{Phaneuf2010} is multiplied by a factor of $10$ to compare it with theory -  $\sigma_{\rm 4d}^{\rm new}$ and
 $\sigma_{\rm 4d}^{\rm old}$ (see text).}

This work was supported by the NSF Grant No.\ PHY-0969386
and a UNA CoA\&S  grant.

}  
\end{document}